\documentstyle[12pt,epsfig,amsmath,amssymb,amsfonts,latexsym]{article}

\makeatletter

\@addtoreset{equation}{section} \makeatother

\newcommand{\be}{\begin{equation}}

\newcommand{\ee}{\end{equation}}

\newcommand{\ba}{\begin{eqnarray}}

\newcommand{\ea}{\end{eqnarray}}

\newcommand{\bea}{\begin{eqnarray*}}

\newcommand{\eea}{\end{eqnarray*}}

\newcommand{\gsim}{\raise.3ex\hbox{$>$\kern-.75em\lower1ex\hbox{$\sim$}}}

\newcommand{\lsim}{\raise.3ex\hbox{$<$\kern-.75em\lower1ex\hbox{$\sim$}}}

\begin{document}

\titlepage

\begin{flushright}
\today \\
\end{flushright}
\vskip 1cm
\begin{center}
{\large \bf Solar Chameleons}
\end{center}

\vspace*{5mm} \noindent

\centerline{ Philippe~Brax$^{}$\footnote{brax@spht.saclay.cea.fr} and Konstantin~Zioutas $^{}$\footnote{zioutas@cern.ch} }

\vskip 0.5cm \centerline{$^{1}$ \em Institut de Physique Th\'eorique}
\centerline{\it CEA, IPhT, CNRS, URA 2306,
  F-91191Gif/Yvette Cedex, France } \vskip 0.5cm
\centerline{$^{2}$\em University of Patras, GR 26504 PATRAS, Greece}

\

\begin{center}
{\bf Abstract}
\end{center}
We analyse the creation of chameleons deep inside the sun and their subsequent conversion to photons near the magnetised surface of the sun.
We find that the spectrum of the regenerated photons lies in the soft X-ray region, hence addressing the solar corona problem.  Moreover, these back-converted photons originating from chameleons have an intrinsic difference with  regenerated photons from axions: their relative polarisations are mutually orthogonal before  Compton interacting with the surrounding plasma. Depending on the photon-chameleon coupling and working in the strong coupling regime of the chameleons to matter, we find that the induced photon flux,  when regenerated  resonantly with the surrounding plasma, coincides with the solar flux within the soft X-ray  energy range. Moreover, using the soft X-ray  solar flux as a prior, we find that with a  strong enough photon-chameleon coupling the chameleons emitted by the sun could  lead to a regenerated photon flux in the CAST pipes, which could  be within the reach of   CAST  with upgraded detector performance. Then, axion helioscopes have thus the potential to detect and identify
particles candidates for the ubiquitous dark energy in the universe.

\newpage
\section{Introduction}

Dark energy is one of the deepest mysteries of present day physics~\cite{Copeland:2006wr,Brax:2009ae}. The current cosmic acceleration~\cite{Riess:1998cb,Perlmutter:1998np} could be explained by the existence
of scalar fields which are ultra-light in the cosmological vacuum. If coupled to matter, these scalar fields would lead to strong violation of the equivalence principle and the detection of a fifth force~\cite{Will:2001mx}. Strong bounds coming from the non-observation of these phenomena would rule out the existence of such scalar fields, if collective phenomena were not at play. One of these mechanisms is the chameleon effect~\cite{Khoury:2003aq,Khoury:2003rn} whereby the coupling of chameleons  to matter leads to a unique density dependent mass for the scalar fields. This is usually enough to screen off the scalar field and evade the fifth force and equivalence principle tests. The coupling of the chameleons to matter almost automatically imply the existence of a coupling to photons~\cite{Brax:2009ey}. The phenomenological consequences of this coupling are numerous for optical cavity experiments~\cite{Upadhye:2009iv}. A similar coupling exists for axions and the axionic production  in the sun could lead to the presence of converted  photons within  confined magnetic flux tubes or more generally the almost ubiquitous large scale solar magnetic fields~\cite{npj}. The solar axions are investigated by axion helioscopes like  CAST~\cite{cast1,cast2,cast3} or Sumico~\cite{sum1,sum2,sum3} providing a strong bound on the inverse coupling $M^{\rm axion}_\gamma \gtrsim 10^{10}$ GeV. Chameleons behave in a completely different manner. Indeed, they are more difficult to produce inside the sun as their mass is density dependent and could exceed  the momentum of the surrounding solar photons in dense regions. This would prevent the production of chameleons by the Primakoff effect. On top of this, chameleons cannot penetrate inside materials where their mass would exceed their energy (energy conservation). As a result, the  CAST apparatus in its present configuration could act as a barrier to the would-be incoming chameleons which could not penetrate inside the experimental device~\cite{pvlas}. However, we will find that the out-streaming chameleons off the sun have an energy in the keV range and can therefore enter the CAST pipes.
Bounds on the chameleon to photon couplings have been obtained using astrophysical observations such as the polarisation of  light~\cite{Burrage:2008ii} and give $M_{\gamma}\ge 10^9$ GeV. On the other hand, the study of the scatter in the luminosity relations of astrophysical objects such as active galactic nuclei~\cite{Burrage:2009mj} tends to favour $M_{\gamma}\le 10^{11}$ GeV. These bounds are valid for chameleon masses $m\le 10^{-12}$ eV in the interstellar medium corresponding to a coupling of the chameleons to matter, which is of order one.

In this letter, we will consider situations where chameleons can be created deep inside the sun thanks to the strong magnetic fields,  which are expected to  necessarily exist,  in order to explain the surface magnetic fields~\cite{tu}.
We will consider chameleon models such that the interstellar mass of the chameleon is larger than $10^{-11}$ eV, which is only achieved in the strong coupling limit of chameleons to matter~\cite{Mota:2006ed,Mota:2006fz}. This  allows one to evade the above  mentioned bounds on $M_\gamma$ and therefore enhance the conversion probability between chameleons and photons. In this case, we find that chameleons can be produced  in the lower convective zone, by conversion of the thermal photons to chameleons inside the macroscopic magnetic fields there. The chameleons can be  produced
by the conversion of  thermal photons  with a spectrum which is thermal-like and is centered in the sub-keV range (300 eV typically). These chameleons turn out to be relativistic and have very little interaction with the sun plasma. In effect we consider them  to have a mean free path larger than the sun radius. On the other hand, the chameleons can be regenerated into  photons by the inverse Primakoff effect. The probability for this to happen is  small but non-negligible. Of course, chameleons could be regenerated into photons inside the convective zone too. The resulting photons would not be observable and would be simply contributing to a tiny change in the radiative transfer of photons inside the sun. On the other hand, regenerated photons in the photosphere, where the mean free path is large,  will escape with a spectrum mimicking the spectrum of the incoming chameleons,  if the photons  Compton scatter only a few times with the surrounding plasma. This process would result in an enhancement of the soft X-ray emission, preferentially above magnetised locations. We find that this increase is most noticeable for smaller values of $M_\gamma$  corresponding to a strong coupling region, which has already been excluded for axions. We also investigate the nature of the chameleon spectrum when they leave the sun and then their potential detection by an axion helioscope like the CAST experiment at CERN. We impose that the number of regenerated photons in the CAST apparatus is below the detection sensitivity of the present experiment. On the whole, the constraint imposed by the recent Sphinx observation of the soft X-ray photon flux from the quiet sun~\cite{syl} and the non-observation of regenerated X-ray  excess photons in the CAST experiment leads to strong restrictions on the parameter space of the chameleon models. More precisely, lowering the value of $M_\gamma$ leads to an over-production of X-ray photons in the CAST pipes while increasing $M_\gamma$ leads to a depletion of the number of regenerated X-ray photons out of the sun. In fact, we find that these two conditions can only be met when the production of photons in the outer sun is further enhanced by a combined resonance-coherence  effect between the out-streaming chameleons and  the surrounding plasma (see~\cite{npj}).

The paper is arranged as follows. In the  first part, we recall some essential facts about chameleons. Then, we describe the chameleon-photon conversion oscillation system and finally we apply it to the situation of interest, which is both the inner and outer sun. Finally, we analyse the constraints imposed by the non-observation of regenerated excess photons from the emitted chameleons in the sun by helioscopes like CAST and the prospects brought forward by an upgraded CAST-like experimental setup~\cite{pap}.

\section{Basic Properties of Solar Chameleons}

Chameleons are scalar fields whose dynamics may explain the origin of the enigmatic cosmic acceleration. Their main feature is a non-trivial coupling to matter, which leads to a density dependent potential, and therefore a mass which is crucially affected by the environment. This is the main feature of these models. As a typical example, dark energy
models postulate the existence of a scalar field $\phi$ whose rolling along a runaway potential $V(\phi)$, i.e. a potential which decreases towards tiny energy scales when the field value is large enough,
would lead to the acceleration of the universe. Unfortunately the mass of the scalar field now is of order
$m_\phi \sim H_0 \sim 10^{-33}$ eV. This is so small that a coupling of the scalar field to matter would lead to violations of Newton's law and the presence of a fifth force.
Fortunately this can be avoided,  if the scalar field couples to gravity in such a way that Newton's constant becomes:
\begin{equation}
G_N(\phi)= e^{2\beta \phi/M_{\rm Pl}}G_N
\end{equation}
where $M_{\rm Pl}\approx 2\cdot 10^{18}$ GeV is the reduced Planck mass. The coupling $\beta$ is a free parameter of the model. We will discuss its value when dealing with the resonance in the photosphere.
In this case, the dynamics of the scalar field are governed by an effective potential in the presence of matter
\begin{equation}
V_{\rm eff}(\phi)= V(\phi) +e^{\beta \phi/M_{ Pl}} \rho
\label{veff}
\end{equation}
where $\rho$ is the matter density surrounding the scalar field.
The
chameleon field fulfills the Klein-Gordon equation with an
effective potential as that of  relation \ref{veff}.
The effective potential has then a minimum obtained solving the equation for $\phi_{\rm min}$, which satisfies the following relation:
\begin{equation}
\partial_\phi V(\phi_{\rm min})= -\beta\kappa_4 e^{\beta \phi_{\rm min }/M_{\rm  Pl}}\rho.
\end{equation}
This is the vacuum of the theory in a given environment.
The  density-dependent minimum is such that  the mass
of the scalar field becomes also density dependent. For this reason, this scalar field is very special and has been dubbed  a chameleon.
The main advantage of this new coupling to gravity resides in the fact that in a dense environment, the mass is such that the range
of the force mediated by the scalar field becomes smaller than 0.1 mm, and it is therefore undetectable. In a sparse environment with very dense matter bodies such as in the solar system,
the chameleon interaction is screened due to another effect called the thin shell mechanism. All in all, chameleons evade the tight gravitational tests and can generate the acceleration of the universe.

We will mainly focus on inverse power law models~\cite{Ratra:1987rm} defined by
\begin{equation}
V(\phi)= \Lambda^4+ \frac{\Lambda^{4+n}}{\phi^n}+ \dots
\end{equation}
where we have neglected higher inverse powers of the chameleon field.
We will choose $\Lambda=2.4~ 10^{-12} {\rm GeV}$ to lead to the acceleration of the universe on large scales.
The potential has a minimum located at
\begin{equation}
\phi_{\rm min}=(\frac{n M_{\rm Pl} \Lambda ^{4+n}}{\beta
\rho})^{1/(n+1)}
\end{equation}
where $\rho$ is the total non-relativistic matter density. The chameleon rest  mass at the minimum is
\begin{equation}
m^2= \beta \frac{e^{\beta \phi_{\rm min}/M_{\rm  Pl}}\rho}{M_{\rm Pl}}\left ( \frac{n+1}{\phi_{\rm min}} + \frac{\beta}{M_{\rm Pl}}\right ).
\label{mass}
\end{equation}
These massive chameleons interact with particles, typically with electrons, due to the coupling given by the Lagrangian
\begin{equation}
{\cal L}_{\rm chameleon-electron}= \beta \frac{m_e}{M_{\rm Pl}} \phi \bar \psi \psi
\end{equation}
Due to the smallness of $m_e\ll M_{\rm Pl}$ and for any reasonable values of $\beta$, the coupling to electrons  is negligible  ($\beta m_e/M_{\rm Pl}\ll 1$). This will imply that
we can neglect the interactions of the chameleons with the whole solar plasma.
We will choose the coupling $\beta$ and the index $n$ in such a way that the mass of the chameleons in the interstellar medium, where the density is around $\rho \approx 10^{-24}~\rm{g/cm^3}$, verifies $m\ge 10^{-11}$ eV.

\section{Chameleonic Primakoff Effect}
The coupling to matter will be generically accompanied with a coupling to photons. In the following, we will consider the coupling to photons as
an arbitrary parameter, which has already been constrained by astrophysical observations.
Hence the chameleons couple to photons with a coupling given by the Lagrangian
\begin{equation}
S_{EM}= -\int d^4 x \sqrt{-g} \frac{e^{\phi/M_\gamma}}{4} F^2
\end{equation}
This form of the coupling has several consequences. The first and most important one is that the chameleon is sensitive to both the matter density and the electromagnetic energy density
\begin{equation}
\rho= \rho_m + \frac{B^2}{2}
\label{dense}
\end{equation}
where $\rho_m$ is the matter density. In this equation, the density is expressed in ${\rm GeV}^4$. For the magnetic part, we use the relation that one Tesla corresponds to $2\cdot 10^{-16} {\rm GeV}^2$. For matter densities we use the conversion $1~\rm{g/cm^3}\approx 4\cdot 10^{-18} {\rm GeV}^4$. For ideal gases at temperature $T$ and pressure $p$ we have $\rho \approx (\frac{ p}{\rm 1 mbar})(\frac{1 K}{ T})\cdot  5\cdot 10^{-23} {\rm GeV}^4$. These allows one to calculate the density of diverse materials and therefore the corresponding mass inside these media.

The chameleon also mixes with photons when  a constant magnetic field is present. This comes from the $\frac{\phi}{M_\gamma }B^2$ coupling in the Lagrangian. Expanding $B= B_0+ b$, where $B_0$ is the constant magnetic field and $b$ the fluctuating field of the photon, we find that the chameleons couple to the polarisation orthogonal to the constant magnetic field. This is to be contrasted with the axion case where the coupling is with the polarisation parallel to the magnetic field. This is a crucial difference in view of distinguishing whether potential signals originate from axions or chameleons. The chameleon mixes with the orthogonal polarisation of the photon, resulting in an effective momentum
\begin{equation}
k^2(\omega )=\omega^2- (m^2- \frac{B^2}{M^2}-\omega^2_{\rm pl})(\frac{\cos \theta +1}{2\cos 2\theta})
\label{k}
\end{equation}
where $\omega$ is the initial frequency of the incoming photons. This depends on
the mixing angle which is given by
\begin{equation}
\tan 2\theta= \frac{2\omega B}{M(m^2-\frac{B^2}{M^2}-\omega^2_{\rm pl})}
\label{theta}
\end{equation}
and the plasma frequency is
\begin{equation}
\omega^2_{\rm pl}= \frac{4\pi \alpha_{EM} n_e}{m_e}
\end{equation}
Electro-neutrality implies that  in the sun $n_e= \frac{\rho_m}{m_p}$, where $m_p$ is the proton mass.
The chameleons propagate when $k^2>0$ and are forbidden to propagate when $k^2<0$. This is what happens in dense materials where
$m^2>\omega^2$ for low values of $\omega$ like in optical cavity experiments, where the energy is in the eV range.
We will find  that $k^2>0$ for chameleons produced deep inside the sun. Moreover we will find
$k\sim \omega$ implying that they are relativistic and therefore can escape from the sun. However, a small fraction can be back-converted into photons preferentially at the magnetised solar photosphere. Such X-rays are  out-streaming off the sun, coming out from magnetic regions while being otherwise unexpected for a cold star like our sun. We will focus on this small fraction of regenerated solar photons.

The thermal photons inside the fully ionised inner sun evolve as free particles for a length equal to their mean free path $\lambda$. Inside the sun, it is an extremely small distance compared to the sun radius. In the solar outer layer, it is much  larger (but still much smaller than the solar radius) and
chameleon production is obtained from the transition probability~\cite{Burrage:2009mj}
\begin{equation}
P_{\rm chameleon}(\omega)= \sin^2 \theta <\sin^2 (\frac{\Delta}{\cos 2\theta})>
\end{equation}
where $m^2_{\rm eff}=m^2-\frac{B^2}{M^2}-\omega^2_{\rm pl}$ and we average out the creation of chameleons over the photon excursion
\begin{equation}
<\sin^2 (\frac{\Delta}{\cos 2\theta})>=\frac{1}{\lambda}\int_0^\lambda dx \sin^2 (\frac{\Delta(x)}{\cos 2\theta})
\end{equation}
where
$\Delta(x)= \frac{m^2_{\rm eff} x}{4\omega }$. As $\Delta (\lambda)\gg 1$ and $\theta \ll 1$ inside the sun, we find that the averaged value is $1/2$ implying that
\begin{equation}
P_{\rm chameleon}(\omega )\approx \frac{1}{2} \theta^2
\label{prob}
\end{equation}
This is the conversion probability of one photon into a chameleon over the length of one mean free path. Notice that the dependence of the conversion probability on the magnetic field and the coupling constant is implicit via relations (\ref{k}) and (\ref{theta}).

\begin{figure}
\begin{center}
\includegraphics[scale=0.5]{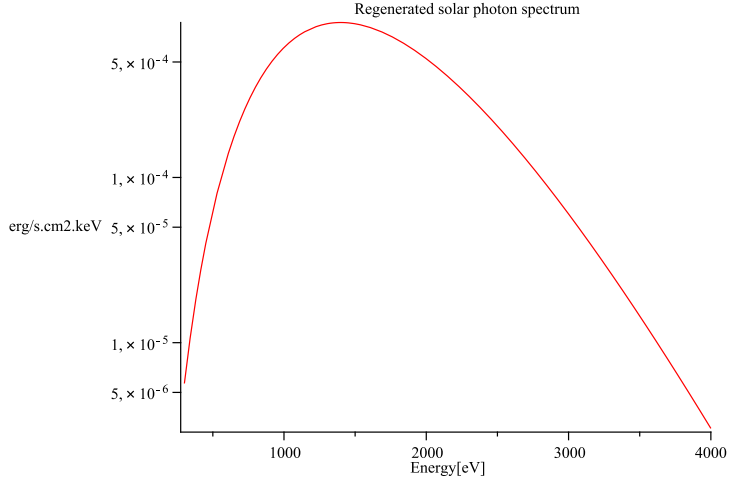}

\caption{The total regenerated photon spectrum. Energies are expressed in eV and the spectrum in erg/s$\cdot\rm cm^2\cdot$keV. The coupling is chosen to be $M_\gamma=10^{5.8}$ GeV and the magnetic field in the convection region is B=30~T. The integrated photon flux at the solar surface  is taken to be $10^{-3}\  \rm{erg\cdot s^{-1} \cdot cm^{-2}}$ for energies greater than $\sim 1 \rm{keV}$, in order to fit the recent Sphinx observation of the quiet sun~\cite{syl}.}
\end{center}
\end{figure}

During one second, the photons experience $N$ interactions inside the rather static solar magnetic field,  where
$
N=\frac{1}{\lambda}
$
in reduced units with $c=\hbar=1$. The probability of creating one chameleon per second out of one thermal photon is then
\begin{equation}
P_{\rm total}(\omega)= NP_{\rm chameleon}(\omega)
\end{equation}
as $P_{\rm chameleon}\ll 1$.
For the thermal photons, we assume  a Planckian distribution
\begin{equation}
p_\gamma(\omega)=\frac{\omega^2}{\pi^2\bar n} \frac{1}{e^{\frac{\omega}{T}}-1}
\end{equation}
where the average number of photons at temperature $T$ is
$
\bar n= \frac{2\zeta(3)}{\pi^2} T^3,
$
implying that the chameleon spectrum  is
\begin{equation}
\Phi_{\rm cham}(\omega)= p_\gamma (\omega) P_{\rm total}(\omega)n_\gamma
\end{equation}
where $n_\gamma$ is the photon flux corresponding to the number of photons going through a sphere of radius R from the centre of the sun.
We take it to be a constant in the magnetic region near the tachocline we are considering.
The total number of chameleons created per second is then
\begin{equation}
N_{\rm cham}=\int_0^\infty \Phi_{\rm cham} d\omega
\end{equation}
The energy spectrum carried away by the chameleons is given by the energy flux
\begin{equation}
l(\omega)= \omega \Phi_{\rm cham}(\omega)
\end{equation}
This leads to a total luminosity
\begin{equation}
L_{\rm cham}=\int_0^\infty l(\omega) d\omega
\label{lum}
\end{equation}
We must impose that this is small compared to the thermal photon luminosity, in order to avoid problems with the solar evolution.

\begin{figure}
\begin{center}
\includegraphics[scale=0.4]{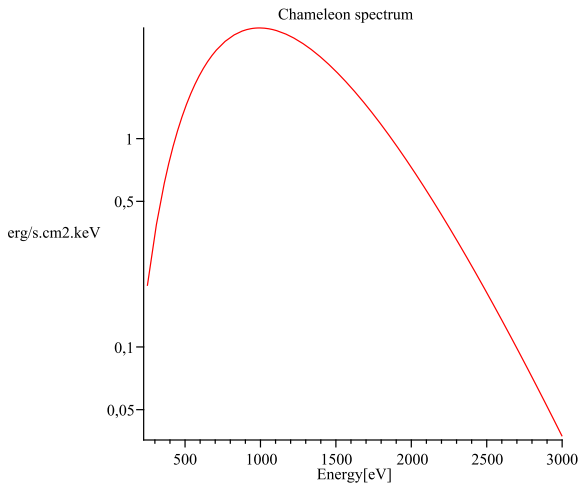}

\caption{The energy spectrum of the emitted chameleons.  Energies are expressed in eV and the spectrum in erg/s$\cdot \rm cm^2\cdot$keV. The coupling is chosen to be $M_\gamma=10^{5.8}$ GeV and  the magnetic field in the lower convection region is B=30~T which is the solar chameleon source. The integrated flux at the solar surface is $\rm{ 4 \ erg\cdot s^{-1}\cdot cm^{-2}}$.}
\end{center}
\end{figure}

Once created, the chameleons interact very little with matter. But, the main source of interaction is the inverse Primakoff effect whereby chameleons can convert to photons. However tiny this photon conversion may well be, it might have some observable effects in the X-ray range of the solar photon spectrum, where the conventional  mechanisms of photon production fail to predict the observed spectrum, and this is the manifestation of the solar coronal heating problem~\cite{npj}. The regenerated photons deep inside the sun are not directly observable and will very quickly thermalise due to the interaction with the surrounding plasma. On the other hand, the photons suddenly created in the photosphere from the chameleon flux  are such that they will not thermalise and escape relatively quickly after their creation with a spectrum reflecting directly the nature of the chameleon production spectrum inside the sun, provided they are not back-converted to X-rays too deep inside the photosphere~\cite{npj}.
The conversion probability to photons is simply
\begin{equation}
P_{\rm photon}(\omega)= \sin^2 \theta <\sin^2 (\frac{\Delta(L_{\rm outer})}{\cos 2\theta})>
\end{equation}
where the averaged value is taken over distances up to $L_{\rm outer}$, the maximal magnetic length over which photons are back-converted from chameleons
\begin{equation}
<\sin^2 (\frac{\Delta(L_{\rm outer})}{\cos 2\theta})>=\frac{1}{L_{\rm outer}}\int_0^{L_{outer}} dx<\sin^2 (\frac{\Delta(x)}{\cos 2\theta})>
\end{equation}
and the parameter $\theta$ is evaluated for values of $B$, $m$ and $\omega_{pl}$ corresponding to the outer sun. The flux of back-converted photons in the photosphere from chameleons created in the tachocline region ($R\sim 0.7 R_{\rm sun}$) is then
\begin{equation}
\Phi_{\rm photon}(\omega)= P_{\rm photon}(\omega)\Phi_{\rm cham}(\omega)
\end{equation}
The spectrum will have a slightly distorted thermal shape as the conversion probability from thermal photons to chameleons is not a flat distribution and increases with the energy $\omega$.

Another important effect is that the light created by the chameleons will be mainly polarised with a linear polarisation perpendicular to the magnetic field in the photosphere. The photon spectrum created by the inverse Primakoff effect will be modified by the Compton interactions with the electrons in the photosphere as considered in~\cite{npj} for the case of solar axions. This will also affect the angular distribution of the photons observed by an outside observer near the earth. All this must be taken into account when considering the exotic solar X-ray emission, either from axions or  chameleons alike. In the following we will concentrate on the back-conversion mechanism of chameleons created in the tachocline region to photons and leave the effect of the interaction with the surrounding plasma for future work.

\section{Phenomenology}

The production of soft X-ray photons in the outer sun by conversion from out-streaming chameleons created inside the sun is constrained by two competing effects. The first one is the Sphinx direct observation of the quiet sun X-ray brightness  specifying that the photon energy flux in the range of energies larger than $\sim 1~\rm keV$ is  about $10^{-3} \rm { erg/s\cdot cm^2}$~\cite{syl}.
Notice that the active sun (corresponding to sun spots) can generate a much larger flux (  $\gtrsim100~ \rm erg/cm^2.s$). In the following we will concentrate on the quiet sun with the same input parameters, although there is a difference between the quiet and the active sun (see discussion in section 6.2 in ~\cite{npj}).  Non-flaring active regions certainly require a modification of the input settings.  Using the quiet sun brightness, we find that
the coupling to photons $M_\gamma^{-1}$ should not be too small. On the other hand, if this coupling is large, we run into a major problem: a large flux of energetic chameleons leaves the sun. These chameleons have an energy in the keV range and therefore would not see matter as a barrier. Hence they would penetrate in helioscopes like CAST or Sumico and get converted into  soft X-ray photons. The CAST experiment gives a tight bound on the number of regenerated photons per hour implying that the coupling must be large enough to prevent an over-production in the magnetic  pipes. Meeting these two constraints imposes a strong restriction on the parameter space of the chameleon models $(M_\gamma, n, \beta)$. In particular, we have observed that the CAST constraint would lead to a negligible photon spectrum out of the sun in the X-ray region, if it were not for a resonance effect with the surrounding plasma near the outer solar layers, which can increase the oscillation length~\cite{npj}. In effect, the mass of the chameleon in the outer sun must be resonantly close to the plasma frequency there, somehow similar to the axion case. This can lead to a fine tuning of the parameters $n$ and $\beta$,increasing thus the chameleon to photon conversion probability (relation \ref{prob}). In the following, we will present a striking example of phenomenology whereby the Sphinx normalisation is achieved while the CAST experiment in its present status could not have seen a photon excess. However, a next generation CAST experiment or even its present configuration with better detector performance  would detect such a low level of X-ray photon excess when pointing at the sun. Of course, it would be of great interest to analyse how generic this result is and carry out a more thorough study of the chameleon parameter space. This is left for future work.

We  focus on chameleon models for which the mass of the chameleon in the interstellar medium is large enough to evade the known bounds on $M_\gamma$. This requires a large coupling of chameleons to matter and we take $M_\gamma= 10^{5.8}$ GeV.
To be specific and to have a strong enough resonance in the outer sun, we choose $\beta=10^{7.09218}$ and $n=8.7$. With these values, the mass of the chameleon in the interstellar medium becomes $m\approx 10^{-5}$eV.

Now, we consider the creation of chameleons in solar regions  where the magnetic field is intense and the  mean free path  small. We focus on  the so-called tachocline zone, a shell of thickness $\delta R \sim 0.05 R_{\rm sun}$ at a radius $R\sim 0.7 R_{\rm sun}$,
which  is assumed to be the source of the solar magnetic field seen at its surface, and which is generally accepted to have a strength of about B = 20-50~T~\cite{tu}, while the density is $\rho_m\approx 0.2 ~{\rm g/cm^3}$. For this work, we take the rather conservative value of B=30~T.
The photon mean free path at this depth is around $\lambda =10$ cm.  The photons have a temperature corresponding to  $\omega_0\approx 200$ eV and the photon flux is around $n_\gamma \approx  10^{21} {\rm s^{-1}cm^{-2}}$. Typically and using  relation \ref{mass},
we find that the chameleons which are created  (in the tachocline region) have a  mass of order $m\approx 18$ eV coming almost entirely from the matter density there (see relation \ref{dense}).

 The tachocline region can  produce a chameleon flux, which can then be converted into photons in the photosphere, where we take $B=0.2$~T and  $\rho_m= 2\cdot 10^{-7}{\rm g/cm^3}$. At the solar surface, we take the average mean free path of the photons  to be below $\lambda
 =10-100$~km, where the out-streaming chameleons are back-converted to photons via the inverse Primakoff effect in the magnetic field of the photosphere, assuming  in this work a full ionization of this region of the photosphere as was argued in \cite{npj}. The typical mass of the chameleons in this outer region is $m\approx 8.9\ {\rm meV} $ almost resonant with the surrounding plasma.

\begin{figure}
\begin{center}
\includegraphics[scale=0.4]{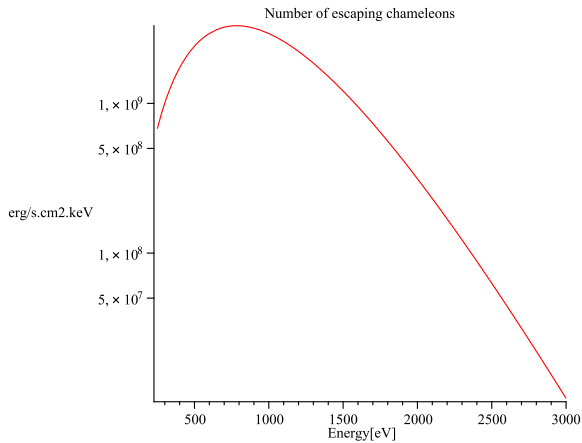}

\caption{The  number of chameleons leaving the sun in $\rm {s^{-1}}\cdot {\rm cm^{-2}}\cdot\rm{ keV^{-1}}$. The integrated number of chameleons at the solar surface is $3\cdot 10^{9}\ \rm{s^{-1}\cdot cm^{-2}}$ assuming the quietest sun brightness of $10^{-3}\rm{erg\cdot s^{-1}\cdot cm^{-2}}$, the magnetic field in the tachocline region B=30~T and the coupling to photons $M_\gamma=10^{5.8}$ GeV.}
\end{center}
\end{figure}

With this we can obtain the regenerated photon flux as the ones obtained by the inverse Primakoff effect from  chameleons emerging from the lower convective zone.
Once a regenerated photon is emitted, it  has to travel a distance of the order of few 100 km below the solar surface before leaving the sun. Before this happens, it performs a random walk
and  escapes  after some scatterings with the surrounding electrons. As a result it completely loses the directionality of the very first regenerated photon as  it was explained in \cite{npj} for the case of the axions. For instance, if the regenerated photons
have to travel a distance $L=1000$~km to leave the sun, they perform a random walk  and escape after several  collisions with the  electrons  $N_\gamma \sim (L/\lambda)^2> 100$. Hence it takes around $t_\gamma\sim 0.03 s$ for the regenerated photons to escape from the sun.

\begin{figure}
\begin{center}
\includegraphics[scale=0.4]{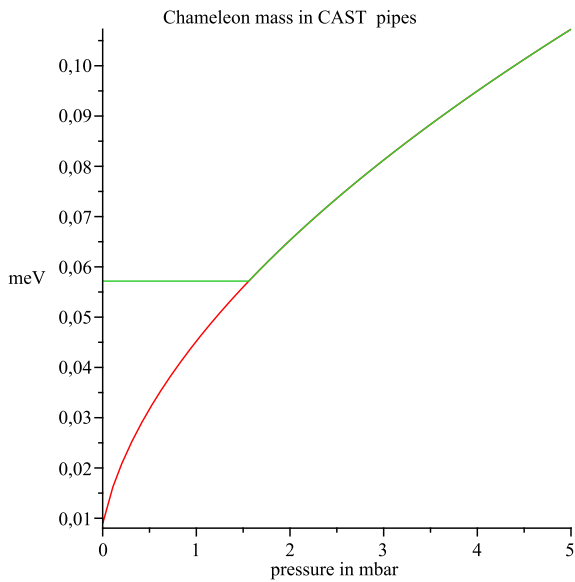}

\caption{The mass of the chameleon in the CAST pipes in meV as a function of the pressure in mbar ($T\approx 1.8 K$). Notice that for values of the pressure smaller than O(2) mbar, the mass is given by $m_{\rm cylinder}$ and remains constant.}
\end{center}
\end{figure}

We have plotted the converted  photon spectrum  taking $M_\gamma=10^{5.8}$ GeV (see figure 1). This gives an integrated photon flux for energies greater than $\sim 1 \rm{keV}$ of order $10^{-3}$ erg/$\rm {s.cm^2}$ corresponding to the result obtained for the quietest  sun by the Sphinx satellite~\cite{syl}. For larger values of $M_\gamma$, the photon flux is suppressed. Let us notice that the escaping  chameleon luminosity ($L_{\rm cham}\approx  4$ erg/$\rm{cm}^2 \cdot\rm s$) is negligible compared to the radiative solar surface brightness   ($L_\gamma \sim 10^{11}$ erg/${\rm cm}^2\cdot\rm s$). We have also presented the spectrum of the chameleon flux out of the sun in figure 2.
These chameleons go through the earth atmosphere as the mass of the chameleons there is $m\approx 0.4$ eV, while the chameleon energy spectrum is peaked in the keV region, resembling actually the directly measured one by the Sphinx mission. Similarly they would enter  the CAST magnet,  since the mass of the chameleons inside all the components of the experimental apparatus,  even  inside  the lead shielding, is $m\lesssim 250$~eV. As we expect solar chameleons of keV energy, they satisfy $k_{\rm chameleon}>m_{\rm device}$ implying that they can get inside the pipes where the 9~T magnetic field is present. There, in the pipes, the chameleons can be back-converted into photons via the inverse Primakoff effect. Now,  the spectrum of the regenerated X-ray photons can be  evaluated using $B_{\rm cast}=9$~T, taking  the length of the magnetic region  to be $L_{\rm cast}=9.26$~m, the diameter of the pipes  $d=43$~mm and the pressure in the vacuum pipes less than $10^{-6}$~mbar ( $T\approx 1.8 ~K$). With these specifications, the mass of the chameleons in the pipes depends on the pressure and the magnetic field and turns out to be  of order $m\approx 5\cdot 10^{-5}$ eV (for the vacuum measurements). More precisely, the mass of a chameleon in a cylinder is given by relation \ref{mass} unless the mass becomes lower than~\cite{optics}
\begin{equation}
m_{\rm cylinder}= \frac{4\sqrt{n+1}}{d}
\end{equation}
This effect springs from the existence of a lower chameleonic  mode with a fixed mass below a continuum where the mass depends on the matter density.
For CAST in the vacuum configuration, the mass is given by $m_{\rm cylinder}$. If the pressure is higher than O(2) mbar, then the mass becomes pressure dependent (see figure 4).

As a result, we can calculate the rate of excess photon production and we find $N_{\gamma}\approx 0.04 $ photon per hour. The CAST experiment has taken data with vacuum in the pipes for $\sim$ 200 hours and the noise level is  0.13 photon per hour in the 1-7 keV band~\cite{cast1,cast2,cast3}. The number of converted photons represents a 1.5 $\sigma$  effect and therefore it could  not have been seen. Though,  a better performing CAST experiment has the potential to observe such an X-ray excess. The situation would improve drastically with a new CAST configuration~\cite{pap}  where we assume  the following specifications:  $B=6 $~T,   $L=15 $~m and  aperture surface $0.15$~$\rm m^2$. In this case, the number of photons per hour becomes $N_\gamma=12$ in the keV region. With a noise level of 4 photons per hour, a 5 $\sigma$ detection would only take 3 hours of solar tracking. Hence we have shown that present and future axion helioscopes have the potential to discover solar chameleons, or, they can provide in parallel  strong constraints on chameleon models. Even if chameleons are not observed in an earth bound helioscope like CAST or Sumico, using as a prior that chameleons are responsible for the soft X-ray  solar flux, we would obtain at least new bounds on the couplings of chameleons to both matter and photons.

\begin{figure}
\begin{center}
\includegraphics[scale=0.4]{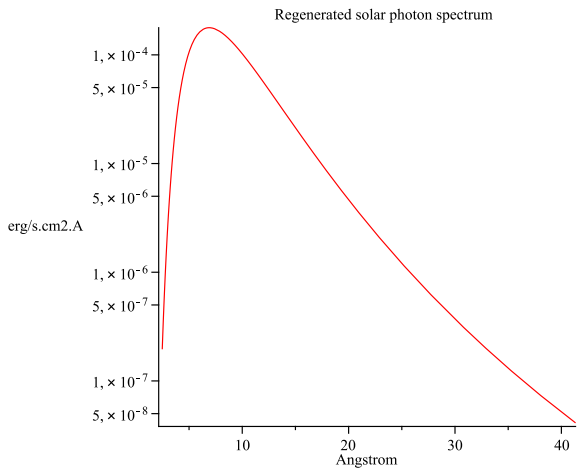}

\caption{The spectrum of regenerated photons out of the sun in $ \rm erg\cdot s^{-1}\cdot cm^{-2}\cdot \AA^{-1}$ as a function of the photon wavelength in Angstrom. The coupling is chosen to be $M_\gamma=10^{5.8}$~GeV and the magnetic field in the regenerating  region at the photosphere is taken to be B=0.2~T.}
\end{center}
\end{figure}

Indeed what we find is that, typically,  the number of regenerated photons in a helioscope goes down as the coupling $M_\gamma^{-1}$ decreases. Using the solar flux in the soft X-ray band as a prior, we find that lowering the couplings require to fine tune the mass of the chameleon in the outer sun closer and closer to a resonance with the local plasma frequency, hence implying a strong correlation between the parameters $n$ and $\beta$.
The solar radial density distribution along with the dynamical behaviour of the sun atmosphere, allow this to happen, occasionally even in an enhanced manner.
Given the potential, i.e. $n$, for the chameleon, and using the solar flux as a prior, the non-observation of regenerated photons by the CAST helioscope while pointing at the sun  and their possible detection in the near future would set at least  limits on the couplings of chameleons to matter $\beta$ and to photons $M_\gamma^{-1}$. This would nicely complement other experimental searches such as the ones in optical (GammeV)~\cite{Upadhye:2009iv} and microwave (ADMX) cavities~\cite{admx}.

\begin{figure}
\begin{center}
\includegraphics[scale=0.4]{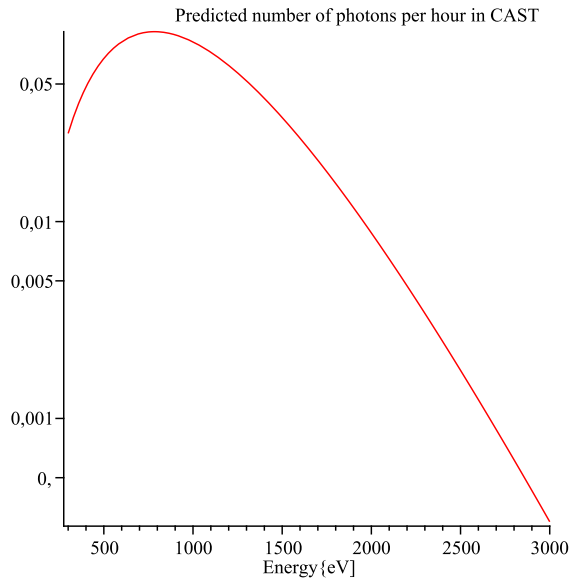}

\caption{The spectrum of regenerated photons as predicted to be seen by   CAST  for one hour of tracking time  and per  $ \rm keV$ as a function of the photon energy in eV.
The pipe length is 9.26 m, the magnetic field is B=9~T and the diameter of the pipe is 43~ mm. This spectrum corresponds to $N_\gamma=0.04$ photon per hour and it makes a 1.5 $\sigma$ effect  above  the background for 200 hours of observation.}
\end{center}
\end{figure}

\section{Conclusion}

We have studied a new mechanism for the production of   solar photons(in the magnetic photosphere) in the  soft X-ray band. We have shown that chameleons produced in the tachocline region of the inner sun, due to the strong magnetic fields there and the coupling of photons to chameleons, can lead to a flux of regenerated photons in the soft X-ray band matching direct solar measurements like that of  the quietest sun by the Sphinx satellite. This results from the resonant regeneration of photons in the outer sun, with the mass of the chameleons closely matching the plasma frequency there. For a given chameleon model, the resonant region is determined by two parameters: the coupling of chameleons to photons $(M_\gamma^{-1})$ and to matter $(\beta)$.
Provided the coupling to photons is not too low, we have shown that the chameleon flux leaving the sun is energetic enough to penetrate axion helioscopes and get converted to X-ray photons in the strong magnetic field present in the magnetic pipes. Taking as template the CAST experiment, we have found that the number of regenerated photons per hour can be within the right ball-park and may be detectable with the upgraded version of the present CAST experiment or a future CAST configuration.
In this work, we have adopted for the CAST experiment a  limit of 0.13 photon per hour over an observation time of 200 hours and for the solar X-ray surface brightness $10^{-3}$ erg/$\rm cm^2 \cdot s$ corresponding to the quietest sun. The mass scale of the photon to chameleon coupling has been chosen to be $M_\gamma =10^{5.8}\rm {GeV}$ allowing one  to deduce that the number of regenerated photons would be 0.04 per hour in the CAST experiment and  would correspond to a 1 $\sigma$ effect for one hour of solar tracking time. On the other hand, the solar surface brightness is associated with uncertainties about the magnetic field strength  in the production zone as well as in the regenerating upper photospheric layers. Nevertheless, we would like to emphasize that our results are indicative of a mechanism which could allow one to understand the mysterious soft X-ray production of the quietest sun and would lead to a foreseeable detection with helioscopes.

In all cases, upgraded helioscopes will provide at least further constraints,  which will allow one to better understand the couplings of chameleons to both matter and photons. The precise determination of the excluded zones in the parameter space due to present and future experimental constraints is certainly interesting and left for future work. The potential existence of magnetic fields in the hot solar core and their consequences for the manifestation of solar chameleons will also (in hard X-rays) be the topic of future work.

\section{Acknowledgments}

We would like to thank A.C.~Davis and D.J.~Shaw for a careful reading of the manuscript and comments. One of us (Ph.~B.) would like to thank the EU Marie Curie Research \& Training network ``UniverseNet" (MRTN-CT-2006-035863) for support. This work was triggered by presentations and discussions during the last two Patras workshops held in DESY (2008) and at the University of Duhram/UK (2009).

\
\end{document}